\documentclass[twocolumn, pra , nofootinbib , superscriptaddress
]{revtex4-1}
\usepackage{graphicx}
\usepackage{amsfonts}
\usepackage{amssymb}
\usepackage{amsthm}
\usepackage{array}
\usepackage{amsmath}
\usepackage{verbatim} 
\usepackage{hyperref}
\usepackage{color}
\usepackage{bbold}
\usepackage{epstopdf}
\usepackage{mathtools}
\usepackage{enumerate}
\usepackage{subcaption}
\usepackage{ulem}
\usepackage{physics}
\usepackage{subcaption}
\usepackage{soul,xcolor}
\usepackage{newtxtext,newtxmath}
\usepackage{hyphenat}

\newtheorem*{proposition*}{Proposition}

\newtheorem*{theorem*}{Theorem}

\newtheorem*{corollary*}{Corollary}

\setstcolor{red}

\begin{document}
\title{Fast quantum imaginary time evolution}
\author{Kok Chuan Tan}
\email{bbtankc@gmail.com}
\affiliation{ School of Physical and Mathematical Sciences, Nanyang Technological University, Singapore 637371, Republic of Singapore}

\begin{abstract}
A fast implementation of the quantum imaginary time evolution (QITE) algorithm called Fast QITE is proposed. The algorithmic cost of QITE typically scales exponentially with the number of particles it nontrivially acts on in each Trotter step. In contrast, a Fast QITE implementation reduces this to only a linear scaling. It is shown that this speed up leads to a quantum advantage when sampling diagonal elements of a matrix exponential, which cannot be achieved using the standard implementation of the QITE algorithm. Finally the cost of implementing Fast QITE for finite temperature simulations is also discussed.
\end{abstract}

\maketitle

\section{Introduction}  

The field of quantum computation has seen fervent activity in the development of algorithmic tools for the computation of the ground states, as well as the thermal states of a physical Hamiltonian $H$\cite{Terhal2000,Bilgin2010, Temme2011, Riera2012, Yung2012, Montanaro2015, Ge2016, Motta2019}. This is in part spurred by the recent availability of noisy intermediate scale quantum (NISQ) hardware\cite{Preskill2018}, which led to many recent practical and numerical demonstrations of the usefulness of quantum computers for simulating many-body physics in the near term\cite{OMalley, Linke2017, Kandala2017, Dumitrescu2018, Klco2018, Colless2018}. At present, such demonstrations do not necessarily outperform classical simulation techniques, but they illustrate the potential of quantum hardware to perform complex computations as the capabilities and scale of quantum hardware continues to improve.

In this note, we consider the recently introduced quantum imaginary time evolution (QITE) algorithm\cite{Motta2019}. The idea behind QITE is to approximate the exponential map $\ket{\psi} \rightarrow e^{H}\ket{\psi}$ up to a normalization factor via a unitary evolution on a quantum computer. Such a map is also called imaginary time evolution\cite{Wick1954} since a unitary evolution $e^{-iHt}$ in real time $t$ generates an exponential map $e^{Ht}$ if the time variable becomes complex, i.e. $t \rightarrow it$. QITE achieves this by breaking down the exponential map $e^{H}$ into smaller Trotter steps, and then approximating each step via a unitary operation by performing a linear optimization of the unitary parameters. QITE has already been successfully tested on current quantum hardware with promising results\cite{Motta2019, Aydeniz2020}.

The cost of implementing a QITE calculation is primarily limited by (i) the number of Trotter steps required to perform a sufficiently accurate approximation, and (ii) the size of the optimization problem required to find the approximate unitary evolution. In particular, the latter can be shown to scale exponentially with the number of particles each Trotter step acts nontrivially on. 

The main result in this note is to demonstrate that this exponential scaling can be brought down to only a linear scaling. This is achieved by showing that any Trotter step that acts nontrivially on $n$ qubits can be reduced to another QITE problem acting nontrivially on $(n-1)$ qubits. This procedure can then be repeated as necessary in order to achieve the required speedup. We call this particular implementation of imaginary time evolution Fast QITE. We then argue that Fast QITE leads to a genuine quantum advantage over known classical algorithms when sampling the diagonal elements of the matrix exponential $e^H$, which standard QITE implementations cannot achieve. Finally, we analyze the cost of implementing Fast QITE for finite temperature simulations, and show that the algorithm is efficient over relevant physical parameters.

\section{Preliminaries}

Consider a general $2^n \times 2^n$ Hermitian matrix $H=\sum_{a} h_a \otimes_{i=1}^n \sigma_{a_i}  = \sum_{a} h_a \sigma_a = \sum_{a} H_a$, where $a_i = 0,1,2,3$, $\sigma_{a_i}$ are Pauli matrices and $a$ is the operator string $a_1 \ldots a_n$. The matrix exponent can be approximated via Trotterization such that 
\begin{align}
	e^{H} = (\prod_{a} e^{\tau H_a })^{1/\tau} + \order{\tau^2}.
\end{align} In this note, each application of $e^{\tau H_a}$ over a small imaginary time step $\tau$ is referred to as a Trotter step.

The QITE algorithm seeks to approximate each imaginary time step $e^{\tau H_a }$ with a unitary operator
\begin{align}
e^{i \tau \sum_{b} x_b \otimes_{i=1}^n \sigma_{b_i}} &= e^{i \tau \sum_{b} x_b \sigma_{b}} \\
&= \prod_b e^{i \tau x_b \sigma_b} + \order{\tau^2}
\end{align} such that the quantity $\norm{e^{\tau H_a}\ket{\psi}/\sqrt{c_a} - e^{i \tau \sum_{b} x_b \sigma_{b}} \ket{\psi}}^2 \approx \norm{e^{\tau H_a}\ket{\psi}/\sqrt{c_a} - (\openone+ i \tau \sum_{b} x_b \sigma_{b})\ket{\psi}}^2$ is minimized. The quantity $c_a \coloneqq \norm{e^{\tau H_a}\ket{\psi}}^2$ is a normalization factor that was introduced because $e^{\tau H_a }$ does not preserve the vector norm. Looking for a solution where  $e^{\tau H_a}\ket{\psi}/\sqrt{c_a} = (\openone+ i \tau \sum_{b} x_b \sigma_{b})\ket{\psi} +\order{\tau^2}$, we can verify that:
\begin{align}
&\norm{e^{\tau H_a}\ket{\psi}/\sqrt{c_a} - (\openone+ i \tau \sum_{b} x_b \sigma_{b})\ket{\psi}}^2 \\
&\approx 2\Im{\tau \sum_{b} \bra{\psi} e^{\tau H_a}x_b \sigma_{b} \ket{\psi} /\sqrt{c_a}} \notag \\
&\quad + 2\Re{\sum_{b,b'} \tau^2 \bra{\psi}x_{b'} x_b\sigma_{b'} \sigma_{b} \ket{\psi}} \\
&= 2 \tau^2 \vec{x}^\dagger (M \vec{x} + \vec{r}),
\end{align}  where\cite{Sun2020}
\begin{align}
M_{bb'} \coloneqq  \Re{ \bra{\psi}\sigma_{b'} \sigma_{b} \ket{\psi}},
\end{align} and 
\begin{align}
r_b \coloneqq \Im{ \bra{\psi} e^{\tau H_a}\sigma_{b} \ket{\psi}} /(\tau \sqrt{c_a}).
\end{align}

 Observe that the above expression is minimized when $\vec{x}$ is the solution to the system of linear equations $M \vec{x} + \vec{r} = 0 $. The Trotter step $e^{\tau H_a}$ is then implemented by sampling the matrix elements of the matrix $M$ and the vector $\vec{r}$. The matrix elements of $M$ can be obtained by checking whether the product of Pauli strings $\sigma_{b'} \sigma_{b}$ is Hermitian or anti-Hermitian, and measuring the expectation value $\bra{\psi}\sigma_{b'} \sigma_{b} \ket{\psi}$ when it is Hermitian (it is zero otherwise). The elements of the vector $\vec{r}$ can be obtained by expanding $\bra{\psi} e^{\tau H_a}\sigma_{b} \ket{\psi}$ and $\sqrt{c_a}$ in terms of $\tau$, which gives us
\begin{align}
&\bra{\psi} e^{\tau H_a}\sigma_{b} \ket{\psi} = \bra{\psi} (\openone + \tau h_a \sigma_a + \tau^2 h_a^2/2 ) \sigma_{b} \ket{\psi} + \order{\tau^3} \\
&c_a^{-1/2} =  \bra{\psi} (\openone - \tau h_a \sigma_a )  \ket{\psi} + \order{\tau^2} \label{eq::normalizationFactor}\\ 
& r_b = h_a \Im{\bra{\psi} \sigma_a \sigma_b  \ket{\psi}} (1- \tau h_a \bra{\psi} \sigma_a  \ket{\psi} ) + \order{\tau^2}.
\end{align}

The cost of implementing QITE for a single Trotter step largely depends on the cost of sampling the matrix elements of $M$. Since $b = b_1 \ldots b_n$, and $b_i = 0,1,2,3$, the total number of elements in $M$ is $\order{2^{2n}}$. For general $H$, we see that this is inefficient as it scales exponentially with system size. The situation can be improved by considering the special case of $k$-local Hamiltonians. In this case, the number of matrix elements that needs to be sampled scales with $\sim 2^{\order{k}}$, which is exponential in $k$, but is manageable for small values of $k$.

\section{Fast QITE}

We now describe a fast implementation of QITE that reduces the exponential scaling of the algorithm down to only a linear scaling. The essential observation here is that standard QITE is expensive to perform because the operator space is acting on too many particles at the same time. For a general $2^n \times 2^n$ Hermitian matrices, the operator space acts on $n$ qubits, resulting in an exponentially scaling complexity. Fast QITE remedies this by providing a systematic method of reducing a QITE problem acting on $n$ qubits, to another QITE problem acting only on $(n-1)$ qubits. This procedure can be performed repeatedly, which dramatically reduces the complexity of the problem.

Consider a Trotter step $e^{\tau \sigma_{a_1}}$ for a single Pauli matrix $ \sigma_{a_1}$. For this imaginary time evolution and input state $\ket{\psi}$, we can find a real time evolution $e^{i\tau \sum_{b_1} x_{b_1} \sigma_{b_1}}$, where $b_1 = 0,1,2,3$ such that up to first order in $\tau$, we have
\begin{align}
e^{i \tau \sum_{b_1} x_{b_1} \sigma_{b_1}} \ket{\psi} \approx e^{\tau \sigma_{a_1}}/\sqrt{c_{a_1}}\ket{\psi}.
\end{align} Expanding both sides to the first order in $\tau$, we obtain
\begin{align}
&(\openone+i\tau \sum_{b_1}x_{b_1} \sigma_{b_1})\ket{\psi} \\
&= (\openone + \tau\sigma_{a_i})(1-\tau \bra{\psi}\sigma_{a_1}\ket{\psi})\ket{\psi} + \order{\tau^2} \\
&=(\openone + \tau\sigma_{a_i}-\tau \bra{\psi}\sigma_{a_1}\ket{\psi})\ket{\psi} + \order{\tau^2},
\end{align} which implies
\begin{align}
i\tau \sum_{b_1}x_{b_1} \sigma_{b_1}\ket{\psi} = (\tau\sigma_{a_1}-\tau \bra{\psi}\sigma_{a_1}\ket{\psi})\ket{\psi} + \order{\tau^2}.
\end{align}

Consider $H_a = h_a \otimes^{n}_{i=1} \sigma_{a_i}$. It can be observed that 
\begin{align}
& i\tau h_a\sum_{b_1}x_{b_1} \sigma_{b_1} \otimes^{n}_{i=2} \sigma_{a_i} \ket{\psi} \\
&= h_a \otimes^{n}_{i=2} \sigma_{a_i} \otimes  (i\tau \sum_{b_1}x_{b_1} \sigma_{b_1}) \ket{\psi} \\
&= h_a \otimes^{n}_{i=2} \sigma_{a_i} \otimes (\tau\sigma_{a_1}-\tau \bra{\psi}\sigma_{a_1}\ket{\psi})\ket{\psi} + \order{\tau^2} \\
&= \tau (H_a -  h_a  \bra{\psi}\sigma_{a_1}\ket{\psi}\otimes^{n}_{i=2} \sigma_{a_i} ) \ket{\psi}  + \order{\tau^2}
\end{align} Reshuffling the terms, we get
\begin{align}
 & \tau h_a  \bra{\psi}\sigma_{a_1}\ket{\psi}\otimes^{n}_{i=2} \sigma_{a_i}  \ket{\psi} \notag \\
 & + i\tau h_a\sum_{b_1}x_{b_1} \sigma_{b_1} \otimes^{n}_{i=2} \sigma_{a_i} \ket{\psi} \\ 
&= \tau H_a \ket{\psi} + \order{\tau^2}.
\end{align} Adding $\openone \ket{\psi}$ on both sides, we obtain up to first order in $\tau$ the approximation
\begin{align}
&e^{\tau h_a  \bra{\psi}\sigma_{a_1}\ket{\psi}\otimes^{n}_{i=2} \sigma_{a_i}} U_{a_1} \ket{\psi} \approx e^{\tau H_a} \ket{\psi}, 
\end{align} where $U_{a_1} \coloneqq \exp{i\tau h_a\sum_{b_1}x_{b_1} \sigma_{b_1} \otimes^{n}_{i=2} \sigma_{a_i}}$. We observe that the original QITE problem involving $n$ qubits acting on the state $\ket{\psi}$ has been reduced to another QITE involving $(n-1)$ qubits acting on the state $U_{a_1} \ket{\psi}$. By letting $H'_a =  h_a  \bra{\psi}\sigma_{a_1}\ket{\psi}\otimes^{n}_{i=2} \sigma_{a_i}$ be the new Hermitian operator in the imaginary time evolution, we can repeat the procedure and continue reducing the number of qubits involved in the QITE. In total, we can iterate this procedure a maximum of $n$ times, until we eventually arrive at
\begin{align}
U_{a_n} \ldots U_{a_1} \ket{\psi} = e^{\tau H_a} \ket{\psi} / \sqrt{c_a}. \label{eq::unitSeries}
\end{align} 

By employing the above reduction process, we see that Fast QITE replaces a single large unitary with a sequence of $n$ smaller unitary operations that has greatly reduced complexity.

\section{Computational Cost of Fast QITE}

We first consider the cost of implementing Fast QITE for a single Trotter step. This is essentially the cost of implementing the series of unitary operations $U_{a_n} \ldots U_{a_1} $ in Eq.~\ref{eq::unitSeries}. An upper bound to this cost is $n$ times the cost of implementing $U_{a_1}$, or equivalently, the cost of performing the Hamiltonian simulation\cite{Berry2015} of $ h_a\sum_{b_1}x_{b_1} \sigma_{b_1} \otimes^{n}_{i=2} \sigma_{a_i}$. This can be achieved using using at most $\order{n}$ number of operations since it is a sum of four products of Pauli operators, each of which performs a maximum of $n$ single qubit operations. The total cost of implementing each Trotter step $e^{\tau H_a}$ is therefore $\order{n^2}$. In comparison to the $\order{2^{2n}}$ scaling of the original QITE algorithm, we see that Fast QITE is exponentially faster for each Trotter step.

For the special case where $H$ is a $k$-local Hamiltonian, each string $a$ consists of at most $\order{k}$ non-identity Pauli matrices. Correspondingly, we only need to perform the reduction process $\order{k}$ times and the cost of performing $U_{a_1}$ is also similarly $\order{k}$. This results in a total cost of $\order{k^2}$ to perform a Trotter step, which is an exponential improvement in terms of $k$ over the scaling of $\order{2^{\order{k}}}$ for standard QITE.

We now consider the full cost of generating the imaginary time evolution $e^H\ket{\psi}/\sqrt{c_H}$, where $c_H \coloneqq \norm{e^{H}\ket{\psi}}^2$. For any arbitrary Hermitian operator $H$, the decomposition $H=\sum_{a} h_a \otimes_{i=1}^n \sigma_{a_i} = \sum_a H_a$ contains up to $2^{2n}$ linearly independent terms. Since Fast QITE implements each Trotter step $e^{\tau H_a}$ using $\order{n^2}$ operations, $\prod_{a} e^{\tau H_a } \ket{\psi}$ can be implemented using a total of $\order{2^{2n} n^2}$ operations. For a fixed precision, the overall cost of generating $e^H\ket{\psi}/\sqrt{c_H}$ is therefore $\order{2^{2n} n^2}$. Since standard QITE requires $\order{2^{2n}}$ operations per Trotter step, it will require $\order{2^{4n}}$ operations to complete the computation. In this case, the advantage of Fast QITE over standard QITE is essentially quadratic.

\section{Sampling diagonal elements of matrix exponentials}

We now compare the performance of Fast QITE to classical algorithms at the task of computing matrix exponentials. Suppose we are interested to sample the diagonal elements of the matrix exponential $e^H$. The most commonly used classical approach for computing these diagonal matrix elements exactly is to diagonalize the matrix such that $H = U D U^\dagger$ where $D$ is a diagonal matrix. The matrix exponential is then computed by first evaluating $e^D$, which can be done efficiently using $\order{n}$ operations, and then evaluating $e^H = U e^{D} U^\dagger$. This diagonalization process can be performed using $\order{2^{3n}}$ operations and the diagonal matrix elements are computed by performing the matrix multiplication $\bra{\psi} U e^{D} U^\dagger\ket{\psi}$. The cost of this classical computation is limited by the diagonalization step, so the overall cost is in this case $\order{2^{3n}}$.

An approximate method with better scaling is to recognize that $e^{tH}\ket{\psi}$ is the solution to the differential equation $d\ket{\psi}/dt = H\ket{\psi}$. We can therefore perform the approximation $\delta{\ket{\psi}} \approx H\ket{\psi} \delta t$ using sufficiently small $\delta t$ a total of $1/\delta t$ times to approximate the evolution $e^{H}\ket{\psi}$. In this approach, each time step requires a matrix-vector multiplication, so the number of operations required is $\order{2^{2n}}$\cite{Moler2003}. In general, we do not expect any classical algorithm to perform faster than $\order{2^n}$, because the vector $e^{H}\ket{\psi}$ has $2^n$ elements, so it requires at least $\order{2^n}$ operations in general to write down.

We show that Fast QITE can be used to sample a diagonal element of $e^H$ with better scaling than classical algorithms. For this purpose, we consider the case where $ H = \sum_{a} H_a$ is a sum of at most $\mathrm{poly}(n)$ terms. Note that under this constraint, each $H_a$ can still act nontrivially on a maximum of $n$ qubits.

We consider the cost of using Fast QITE to compute the diagonal element. This can be achieved by performing a projection of $e^{H}\ket{\psi}/\sqrt{c_H}$ onto the state $\ket{\psi}$ with probability  $ \abs{\bra{\psi }e^{H}\ket{\psi}}^2/c_H$. The normalization factor $c_H$ can be obtained by observing that each Trotter step generates a normalized state $e^{\tau H_a }\ket{\psi}/\sqrt{c_a}$, so $c_H$ can be obtained from the the multiplication of all the normalization factors $c_a$ obtained at each Trotter step (see Eq.~\ref*{eq::normalizationFactor}). Since there are $\mathrm{poly}(n)$ Trotter steps in total, finding $c_H$ requires $\order{\mathrm{poly}(n)}$ number of operations. $\order{\mathrm{poly}(n) n^2} = \mathrm{poly}(n)$ is the cost of generating the state $e^{H}\ket{\psi}/\sqrt{c_H}$ using Fast QITE, so the overall cost of sampling the diagonal element can be done in polynomial time. 

In comparison, standard QITE will require $\order{2^{2n} \mathrm{poly}(n)}$ operations to perform a similar computation, which is slower than a classical algorithm employing the differential equation method. Even if we account for the fact that $H$ is $\mathrm{poly}(n)$ sparse, the classical differential equation method can do no better than $\order{2^n \mathrm{poly}(n)}$ scaling due to the fact that $e^{H}\ket{\psi}$ has $2^n$ vector elements.  Fast QITE is therefore exponentially faster at this task than both classical and standard QITE algorithms.

\section{Finite temperature simulation with Fast QITE}

In the previous section, we considered the use of Fast QITE to sample the diagonal elements of the matrix exponential $e^H$. This problem is of particular interest because of its immediate application in the simulation of many body systems at finite temperature. Here, we perform a cost analysis of using Fast QITE to simulate a system in thermal equilibrium with a heat bath at inverse temperature $\beta$. For such systems, the density matrix of the system takes on the form $\rho = e^{-\beta H}/Z$ where $H$ is typically assumed to be some $k$-local Hamiltonian and $Z= \tr(e^{-\beta H})$ is the partition function. The goal is to measure an observable $O$ such that $\expval{O}_{\rho} = \tr(O\rho )$. We will initially consider the case where the eigenvalues and eigenvectors of $O$ are known. 

It is not difficult to see that Fast QITE can useful under such conditions. Let $\{ \ket{i} \}_{i=1}^{2^n}$ be the eigenbasis of the Hermitian observable $O$. We observe that 
\begin{align}
\expval{O}_{\rho} &= \tr(O\rho ) \\
&=\sum_{i =1}^{2^n} \bra{i} O \rho \ket{i} \\
&=\sum_{i =1}^{2^n} \bra{i} O \ketbra{i} e^{-\beta H} \ket{i} /Z .
\end{align} Since $ \sum_{i=1}^{2^n}\ketbra{i} e^{-\beta H} \ket{i} /Z =1$, we can define the probability distribution $p_i \coloneqq \bra{i} e^{-\beta H} \ket{i} /Z$ and write the expectation value as the statistical average
\begin{align}
\expval{O}_{\rho} = \sum_i \bra{i} O \ket{i} p_i.
\end{align} The thermal statistical average can therefore be obtained by sampling from the eigenvalues of $O$ with probability $p_i$. The distribution $p_i$ can be simulated using standard Monte Carlo techniques so long as $\bra{i} e^{-\beta H} \ket{i}$ is computable. We see that this is just the diagonal element of the matrix exponential, so it can be sampled using Fast QITE. Due to the assumption of $k$-locality, the number of Trotter steps required to is $\order{\beta n k}$, each of which requires $\order{k^2}$ operations in Fast QITE. The total cost is therefore $\order{\beta n k^3}$ to obtain each Monte Carlo sample using Fast QITE. Standard QITE will achieve this at the cost of $\order{\beta n 2^{\order{k}}k}$ , so the speedup here is exponential in terms of $k$.

In cases where the eigenvalues and eigenvectors of $O$ cannot be efficiently computed (for instance, when $O$ is the Hamiltonian $H$ of the system itself), a version of the  minimally entangled typical thermal states (METTS) method\cite{Stoudenmire2010} may be performed using QITE\cite{Motta2019}. Let $\{ \ket{i} \}_{i=1}^{2^n}$ be any complete orthogonal basis. The METTS method rewrites the thermal statistical average in the following way:
\begin{align}
\expval{O}_{\rho} &= \tr(O\rho ) \\
&= \tr(O e^{-\beta H})/Z \\
&= \tr( e^{-\beta H/2} O e^{-\beta H/2})/Z\\
&= \sum_{i} \bra{i}e^{-\beta H/2} O e^{-\beta H/2}\ket{i}/Z \\
&=  \sum_{i} \frac{\bra{i}e^{-\beta H/2} O e^{-\beta H/2}\ket{i}}{w_i} \frac{w_i}{Z} \\
&= \sum_{i} \frac{\bra{i}e^{-\beta H/2} O e^{-\beta H/2}\ket{i}}{w_i} p_i,
\end{align} where $w_i \coloneqq \bra{i}e^{-\beta H}\ket{i}$ and $p_i = w_i/Z$. We can verify that $\sum_{i=1}^{2^n} p_i = 1$, so once again, we can apply standard Monte Carlo techniques to obtain the statistical average, so long as we are able to compute $w_i$ and $\bra{i}e^{-\beta H/2} O e^{-\beta H/2}\ket{i}$. $w_i$ is just a diagonal element of a matrix exponential, which can be obtained using Fast QITE with $\order{\beta n k^3}$ operations.  $\bra{i}e^{-\beta H/2} O e^{-\beta H/2}\ket{i}$ can be obtained by generating a state $ \propto e^{-\beta H/2}\ket{i}$ using Fast QITE and then measuring the observable $O$ on this state. If this measurement can be performed efficiently, then $\bra{i}e^{-\beta H/2} O e^{-\beta H/2} \ket{i}$ can also be efficiently sampled. Suppose we are trying to measure the system energy, so $O = H$. In this case $O$ is $k$-local so we can measure $O$ using at most $\order{nk}$ operations. Performing Fast QITE to generate a state $\propto e^{-\beta H/2}\ket{i}$ costs $\order{\beta n k^3}$ so the overall cost of performing each Monte Carlo step is $\order{\beta n^2k^4}$.

\section{Conclusion}
In this note, we proposed a method of reducing the complexity of QITE. The resulting algorithm has better scaling properties when compared to the standard implementation of the QITE algorithm. For this reason, we call the new algorithm Fast QITE. This reduction in complexity is achieved by iteratively reducing a QITE problem involving $n$ qubits, to another QITE problem acting only on $(n-1)$ qubits. As the operator space grows exponentially in size $n$, this has the effect of significantly reducing the dimensionality of the operator space, leading to cost savings. 

For an arbitrary Hermitian operator $H = \sum_a H_a$, each Trotter step $e^{\tau H_a}$ can be implemented in Fast QITE using only $\order{n^2}$ operations. That is an exponential improvement compared to the $\order{2^{2n}}$ scaling of the standard QITE implementation. 

We then analyzed the cost of using Fast QITE to perform certain tasks related to matrix exponentiation. The first is to sample the diagonal elements of the matrix $e^{H}$. Here, we argued that there is a quantum advantage over classical algorithms. For the special case where $H = \sum_a H_a$ is a sum of at most $H = \sum_a H_a$ terms, the cost of sampling the diagonal element of $e^H$ costs $\order{\mathrm{poly}(n)}$ operations using Fast QITE. This is compared to the $\order{2^{2n}\mathrm{poly}(n)}$ scaling of standard QITE. The cost of a classical algorithm is also not less than $\order{2^n}$. The cost savings for the full imaginary time evolution is therefore exponential in terms of the number of qubits $n$ using Fast QITE when compared to both classical and standard QITE algorithms. For the special case where $H$ is a $k$-local Hamiltonian, the cost savings is exponential in terms of $k$.

We also considered the finite temperature simulation of a system with $k$-local Hamiltonian $H$, at inverse temperature $\beta$. The goal here is to find the thermal average of an observable $O$ such that $\expval{O}_{\rho} = \tr(O\rho)$ where $\rho$ is a state in thermal equilibrium. Fast QITE can be combined with standard Monte Carlo techniques to calculate the statistical average. For the case where the eigenvalues and eigenvectors of $O$ are known, we show that each Monte Carlo step can be performed using $\order{\beta n k}$ operations. In the case where the eigenvalues and eigenvectors are not known but $O$ is $k$-local (such as when when $O$ is the system Hamiltonian),  the cost of each Monte Carlo step is $\order{\beta n^2 k^4}$. In both cases, Fast QITE scales linearly with the inverse temperature $\beta$, and polynomially with the number of qubits $n$ and number of nearest neighbours $k$, which demonstrates the efficiency of Fast QITE with respect to the relevant parameters $\beta$, $n$ and $k$ for finite temperature simulation.

\acknowledgments K.C. Tan was supported by the NTU Presidential Postdoctoral Fellowship program funded by Nanyang Technological University.

\end{document}